\documentclass[reprint,
superscriptaddress,
 amsmath,amssymb,
 aps,
pra,
]{revtex4-1}

\usepackage{graphicx}
\usepackage{dcolumn}
\usepackage{bm}

\begin{document}


\title{Four-wave mixing parametric oscillation and frequency comb generation at visible wavelengths in a silica microbubble resonator}
\affiliation{Light-Matter Interactions Unit, Okinawa Institute of Science and Technology Graduate University, Onna, Okinawa 904-0495, Japan}
\affiliation{Department of Electrical and Systems Engineering, Washington University, St. Louis, Missouri 63130, USA}
\affiliation{National Engineering Laboratory for Fiber Optics Sensing Technology, Wuhan University of Technology, Wuhan, 430070, China}
\author{Yong Yang}
\affiliation{Light-Matter Interactions Unit, Okinawa Institute of Science and Technology Graduate University, Onna, Okinawa 904-0495, Japan}
\affiliation{National Engineering Laboratory for Fiber Optics Sensing Technology, Wuhan University of Technology, Wuhan, 430070, China}
  \email{yong.yang@oist.jp}

\author{Xuefeng Jiang}%

\affiliation{Department of Electrical and Systems Engineering, Washington University, St. Louis, Missouri 63130, USA}%


\author{Sho Kasumie}
\affiliation{Light-Matter Interactions Unit, Okinawa Institute of Science and Technology Graduate University, Onna, Okinawa 904-0495, Japan}%
\author{Guangming Zhao}
\affiliation{Department of Electrical and Systems Engineering, Washington University, St. Louis, Missouri 63130, USA}
\author{Linhua Xu}
\affiliation{Department of Electrical and Systems Engineering, Washington University, St. Louis, Missouri 63130, USA}
\author{Jonathan Ward}
\affiliation{Light-Matter Interactions Unit, Okinawa Institute of Science and Technology Graduate University, Onna, Okinawa 904-0495, Japan}%
\author{Lan Yang}
\affiliation{Department of Electrical and Systems Engineering, Washington University, St. Louis, Missouri 63130, USA}%
\author{S\'{i}le Nic Chormaic}
\affiliation{Light-Matter Interactions Unit, Okinawa Institute of Science and Technology Graduate University, Onna, Okinawa 904-0495, Japan}%

\date{\today}

\begin{abstract}
Frequency comb generation in microresonators at visible wavelengths has found applications in a variety of areas such as metrology, sensing, and imaging. To achieve Kerr combs based on four-wave mixing in a microresonator, dispersion must be in the anomalous regime. In this work, we  demonstrate dispersion engineering in a microbubble resonator (MBR) fabricated by a two-CO$_2$ laser beam technique. By decreasing the wall thickness of the MBR down to 1.4 $\mu$m, the zero dispersion  wavelength shifts to values shorter than 764 nm, making phase matching possible around 765 nm. With the optical \textit{Q}-factor of the MBR modes being greater than $10^7$, four-wave mixing is observed at 765 nm for a pump power of 3 mW. By increasing the pump power, parametric oscillation is achieved, and a frequency comb with 14 comb lines is generated at visible wavelengths.
\end{abstract}

\maketitle

A frequency comb is a light source with equidistant lines in its optical spectrum. It can be generated from four-wave mixing (FWM) and mediated by hyper-parametric oscillation \cite{Udem2002}. In modern optics, frequency combs have applications in many areas, such as frequency metrology \cite{Udem2002}, precise optical clocks \cite{Kippenberg2011}, sensing, and biomedical imagining \cite{Fercher2003}. In the last decade, whispering gallery mode resonators (WGMRs) have emerged as excellent devices for frequency comb generation. Compared to standard fiber-based frequency combs\cite{Cundiff2003}, WGMR-based combs are miniature in size and do not require a high-power fs laser to drive the comb - these benefits arise from their ultra-high \textit{Q}-factor and small mode volume. Frequency combs have been realized near the telecommunications bands in different types of WGMRs, such as microspheres \cite{Agha2009}, microtoroids \cite{DelHaye2007,DelHaye2011}, microrings \cite{Okawachi2011}, microdisks \cite{Savchenkov2008,Grudinin2009,Kippenberg2011science}, and microbubbles \cite{Li2013,Farnesi2015,Yang2016,Lu2016}.

Frequency combs in WGMRs require phase matching over a broadband frequency range where the group velocity dispersion (GVD) plays an important role and is crucial for achieving the maximum comb bandwidth \cite{Agha2007}. In WGMRs, the GVD is determined by (i) the material dispersion described by the Sellmeier formula and (ii) the geometric dispersion due to non-equidistant mode distribution in the resonator. By changing the material and selecting higher-order whispering gallery modes, the zero dispersion wavelength (ZDW) can be shifted towards longer wavelengths \cite{Lin15,Riesen2015,Lin2015}, thus expanding the possibility of frequency comb generation to the mid-IR range \cite{Wang2013,Luke2015,Savchenkov2015}

Apart from extending the wavelength range of a frequency comb towards the infrared region of the spectrum, it is also highly desirable to do the opposite, \textit{i.e.}, moving the frequency comb to shorter wavelengths,including the visible region.  For example, a frequency comb around 780 nm can be used to lock a laser to the rubidium $D_2$ transitions \cite{Vanier2005}, as required in atomic clocks. Moreover, in a water environment, light from a near-infrared (NIR) or mid-infrared frequency comb will be strongly absorbed, whereas the absorption of red or near-red (such as 780 nm) light is much less.  Hence, a frequency comb at 780 nm could also be used for such purposes as biological sensing and optical computed tomography (OCT) imaging \cite{Fercher2003,Swanson1992}. To obtain a Kerr frequency comb, the ZDW must be shifted towards the visible range. However, this is challenging because of the material dispersion.

To date, three methods for realizing frequency combs in the visible spectral region have been reported.  A silicon nitrate microring was used to generate a NIR comb around 1540 nm, which was converted to the near visible range by exploiting the material's strong second-order optical nonlinearity \cite{Miller2014}. The authors observed 17 comb lines in the 765-775 nm region. In the second method,  assisted by a mode-interaction process, a frequency comb in the anomalous dispersion regime was used to generate new combs in the normal dispersion regime \cite{Xue2015}. Hence, it may be exploited to generate a comb in the visible range. The third technique relies on engineering the total dispersion of the system.  This can be achieved by out-of-plane excitation of the higher order bottle modes in WGMRs with a parabolic lateral profile \cite{Savchenkov2011}. In such a situation, the high \textit{Q }optical modes propagate along the axis of symmetry. It has been shown that the geometry dispersion of such modes is strongly related to the lateral profile \cite{Savchenkov2011,Yang2016}. By choosing the right resonator profile, even at a center wavelength of 780 nm, the total dispersion can be forced into the anomalous regime. Using this technique, a frequency comb centered at 794 nm was observed in a crystalline WGMR \cite{Savchenkov2011}. Alternatively, by carefully designing the lateral profile of a wedged silica microdisk, broadband dispersion control can be obtained \cite{YangKiYoul2016}. Such advances indicate that visible range frequency combs can be achieved.

In this Letter, we provide a much more controllable method to engineer the dispersion by using a microbubble resonator (MBR). An MBR is a hollow structure with a spherical outer profile\cite{White2006,Sumetsky2010,Ward2013,Yang:16}. In 2013, FWM parametric oscillation at the telecommunications wavelength around 1550 nm was first reported in an MBR \cite{Li2013} and, later, frequency comb generation was also realized \cite{Farnesi2015,Yang2016,Lu2016}. In an air-filled silica MBR, the mode is distributed both in the wall and in the inner air owing to the evanescent field penetration. By varying the wall thickness, the proportion of the mode intensity in the air can be modified, thus changing the effective index and the frequency distribution of cavity modes \cite{Yang2014}. It has been theoretically shown that, by shrinking the wall thickness, the ZDW shifts towards shorter wavelengths \cite{Li2013,Riesen:16}. Therefore, it should be possible to generate a frequency comb in the visible range by optimizing the wall thickness of the MBR. In the following, details of how to experimentally achieve a Kerr frequency comb which extends from 758-775 nm using an MBR is described.
\begin{figure}
\centering
 \includegraphics[width=.8\columnwidth]{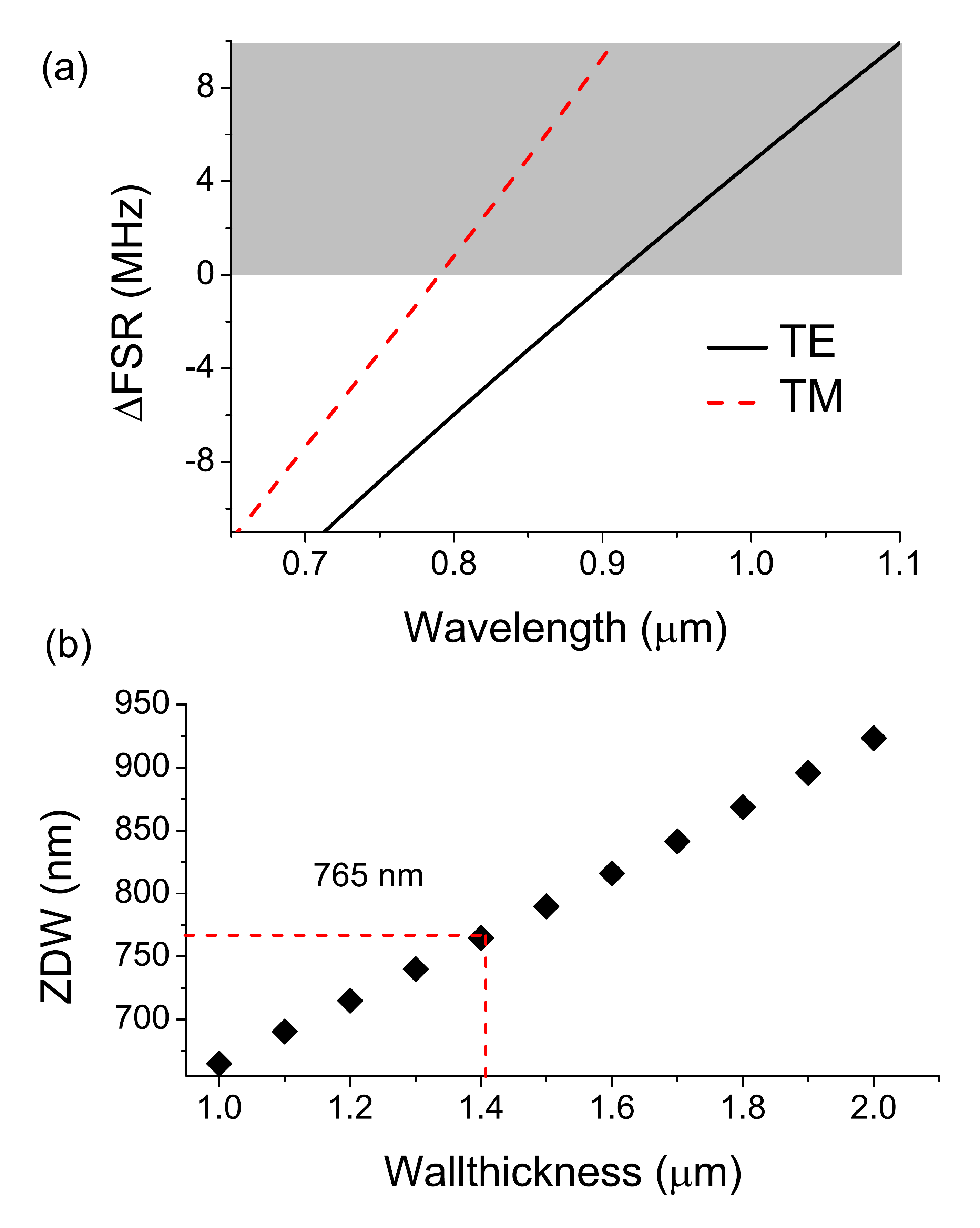}
 \caption{\label{fig:cal}(a) The total dispersion (in terms of derivation of the FSR) of WGMs in a microbubble with wall thickness 1.5 $\mu$m. The grey area is the anomalous dispersion region.   (b) The zero dispersion wavelength of a microbubble with wall thickness varying from 1.0 $\mu$m to 2.0 $\mu$m.  $R=60$ $\mu$m in (a) and (b).}
\end{figure}

A numerical method was used to find the eigenfrequency of an MBR with outer and inner radii, $R$ and $r$, respectively \cite{Riesen2015}. This is done by solving the characteristic equation \cite{Meldrum2014} as below:
\begin{equation}
\begin{aligned}
& \frac{n_0^pH^{(1)}_l{}'(n_0kR)}{n_s^p H_l^{(1)}(n_0kR)}=\frac{B_lH_l^{(2)}{}'(n_skR)+H_l^{(1)}{}'(n_skR)}{B_lH_l^{(2)}(n_skR)+H_l^{(1)}(n_skR)}.\\
& B_l=\frac{(n_s/n_0)^pJ_l(n_0kr)H_l^{(1)}{}'(n_skr)-J_l'(n_0kr)H_l^{(1)}(n_skr)}{-(n_s/n_0)^pJ_l(n_0kr)H_l^{(2)}{}'(n_skr)+J_l'(n_0kr)H_l^{(2)}(n_skr)}.
\end{aligned}
\label{eq:EVE}
\end{equation}
Here, $p$ is the mode number for polarization ($p=1$ for TE and $p=-1$ for TM), $l$ is the polar mode number, $n_0$ and $n_s$ are the refractive indices of air and the bubble shell (silica in our case), respectively. $J_l$ and $H_l^{(1,2)}$ are Bessel and Hankel functions of the first and second kind, respectively. The total dispersion is calculated once the eigenfrequency (\textit{i.e.}, the wave number) of the bubble mode, $k$, is obtained from Eq. \ref{eq:EVE} by evaluating the variation of the free spectral range (FSR). In Fig. \ref{fig:cal}(a), we show the eigenfrequencies of an MBR with a diameter of 120 $\mu$m and a wall thickness of 1.5 $\mu$m. TE mode and TM modes have different dispersions, as explained in \cite{Riesen2016}. We see that, for the TM mode, the ZDW is about 790 nm. The calculated ZDW for different wall thicknesses is plotted in Fig.\ref{fig:cal}(b).  To shift the ZDW to 765 nm, the wall thickness should be reduced to around 1.4 $\mu$m. The threshold for FWM depends on $Q^2/V$, where $V$ is the WGM  volume. \textit{Q} is lower for a thinner MBR; therefore, its geometrical parameters should be optimized \cite{Riesen:16}. In practice, obtaining an MBR with a high enough \textit{Q}-factor was a major technical challenge until an improved fabrication process was developed \cite{Yang:16}.

\begin{figure}
\centering
\includegraphics[width=.9\columnwidth]{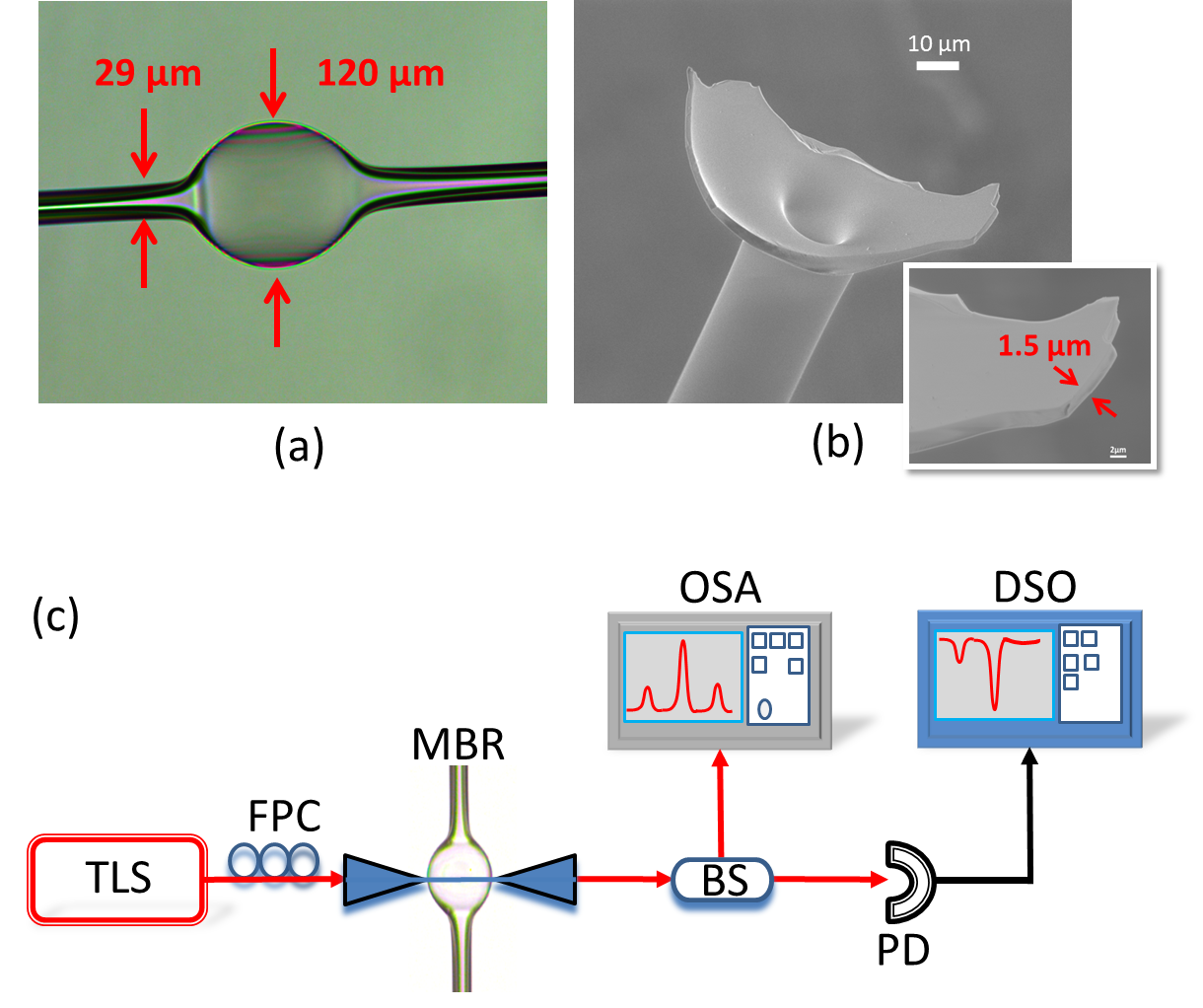}
\caption{\label{fig:setup} (a) A microscope image of the MBR. The diameter is measured to be 120 $\mu$m. (b) An SEM image showing the cross-section of the MBR. The typical wall thickness is 1.5 $\mu$m as shown in the inset. The actual wall thickness should be less than this, as explained in the main text. (c) The experimental setup for measuring the frequency comb in the MBR. TLS: tunable laser source; BS: beam splitter; FPC: fiber polarization controller; PR: photoreceiver; OSA: optical spectrum analyzer; DSO: digital storage oscilloscope. }
\end{figure}

In our experiment,  using the two CO$_2$ laser beam setup described previously \cite{Yang:16},  an ultrahigh \textit{Q}-factor microbubble, with a wall thickness less than 1 $\mu$m can be fabricated. To obtain the desired wall thickness of around 1.4 $\mu$m, we chose a silica capillary with an inner:outer diameter of 100 $\mu$m:375 $\mu$m, which we subsequently tapered down to an outer diameter of 29 $\mu$m by a heat-and-pull technique. The tapered microcapillary was then connected to a dry nitrogen gas with an aerostatic pressure of about 2.5 bar. By heating the microcapillary on both sides using the CO$_2$ laser beams, its wall softens and swells, and the process can be monitored on  a CCD camera. When the outer diameter of the bubble reached about 120 $\mu$m, the process was stopped. The geometric parameters were estimated under a microscope, see Fig. \ref{fig:setup}(a). A more precise measurement of the wall thickness was implemented after the experiment by breaking the microbubble near the middle and performing a scanning electron microscopy (SEM). From Fig. \ref{fig:setup}(b), the wall thickness was estimated to be about 1.5 $\mu$m. Note that the cross-section in Fig. \ref{fig:setup}(b) is slightly removed from the equatorial plane of the MBR, and, therefore, the actual wall thickness is probably thinner than 1.5 $\mu$m.

A tapered fiber was used to couple light into/out of the MBR; single mode fiber (in the wavelength bandwidth of 780 nm) was tapered down to a  diameter of less than 1 $\mu$m. The total light propagation efficiency of the taper was about 70\%. The MBR was placed on a 3D nano-stage to control the coupling between the resonator and the tapered fiber.   In order to excite and measure FWM and the frequency comb generated in the MBR, a setup as illustrated in Fig. \ref{fig:setup}(c) was used. Pump power from a tunable laser diode (New Focus TLB-6712-P), centered at 775 nm, was coupled into the MBR through the tapered fiber. A maximum power of 6 mW could be coupled into the MBR. After the light coupled out of the MBR, it  passed through a 50/50 inline beam splitter (BS). One output of the BS was connected to a photoreceiver (New Focus 1801)
and the other one was connected to an optical spectrum analyzer (OSA, HP 70950B) with a minimum resolution of 0.08 nm.

\begin{figure}
\centering
\includegraphics[width=.9\columnwidth]{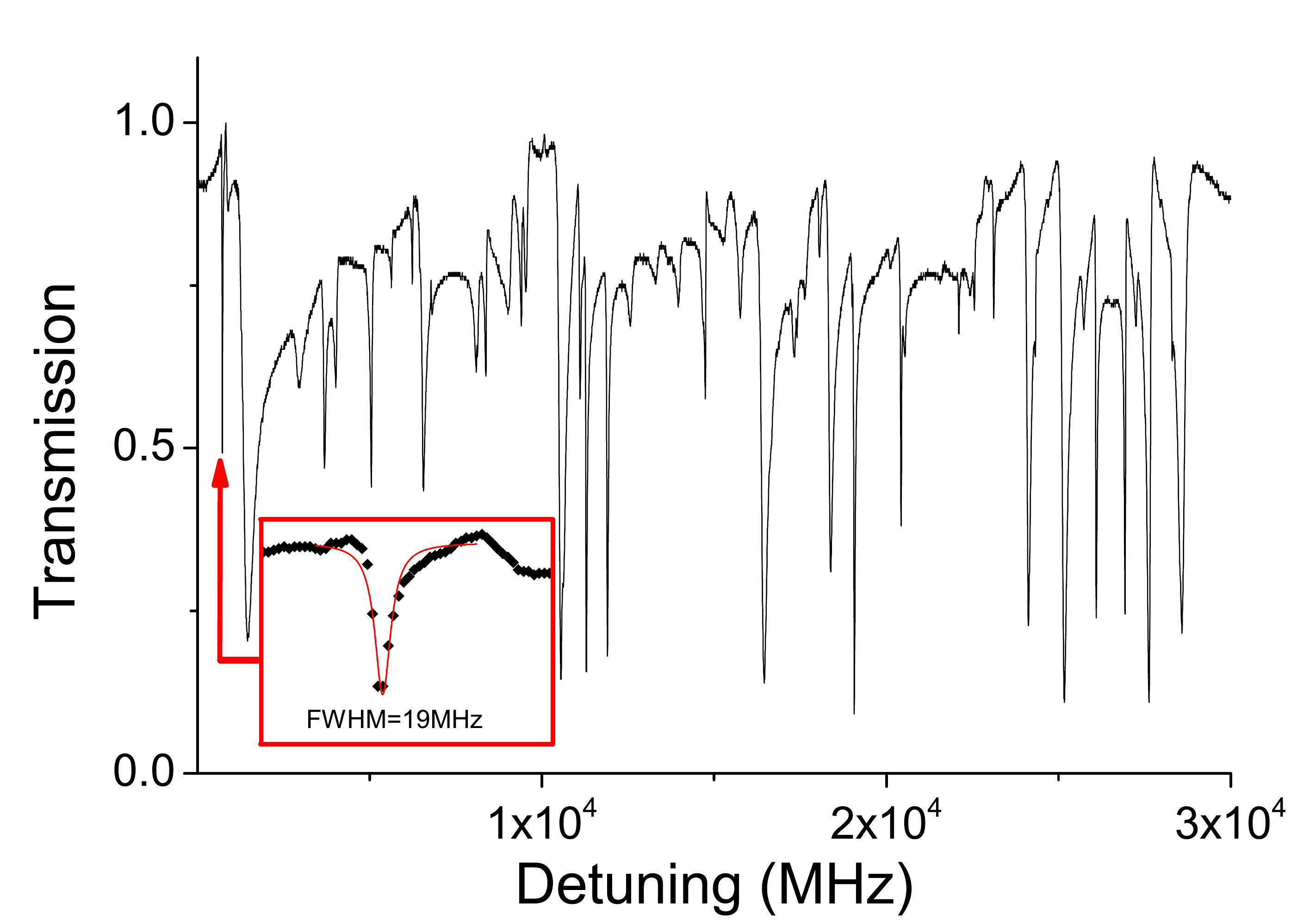}
\caption{\label{fig:trans} The transmission spectrum of the MBR. The inset shows the transmission of the pump WGM in the experiment. The red curve is a Lorentzian fit with a full-width-at-half-maximum of about 19 MHz, corresponding to a \textit{Q}-factor of about $2\times 10^7$. }
\end{figure}
\begin{figure}
\centering
\includegraphics[width=.9\columnwidth]{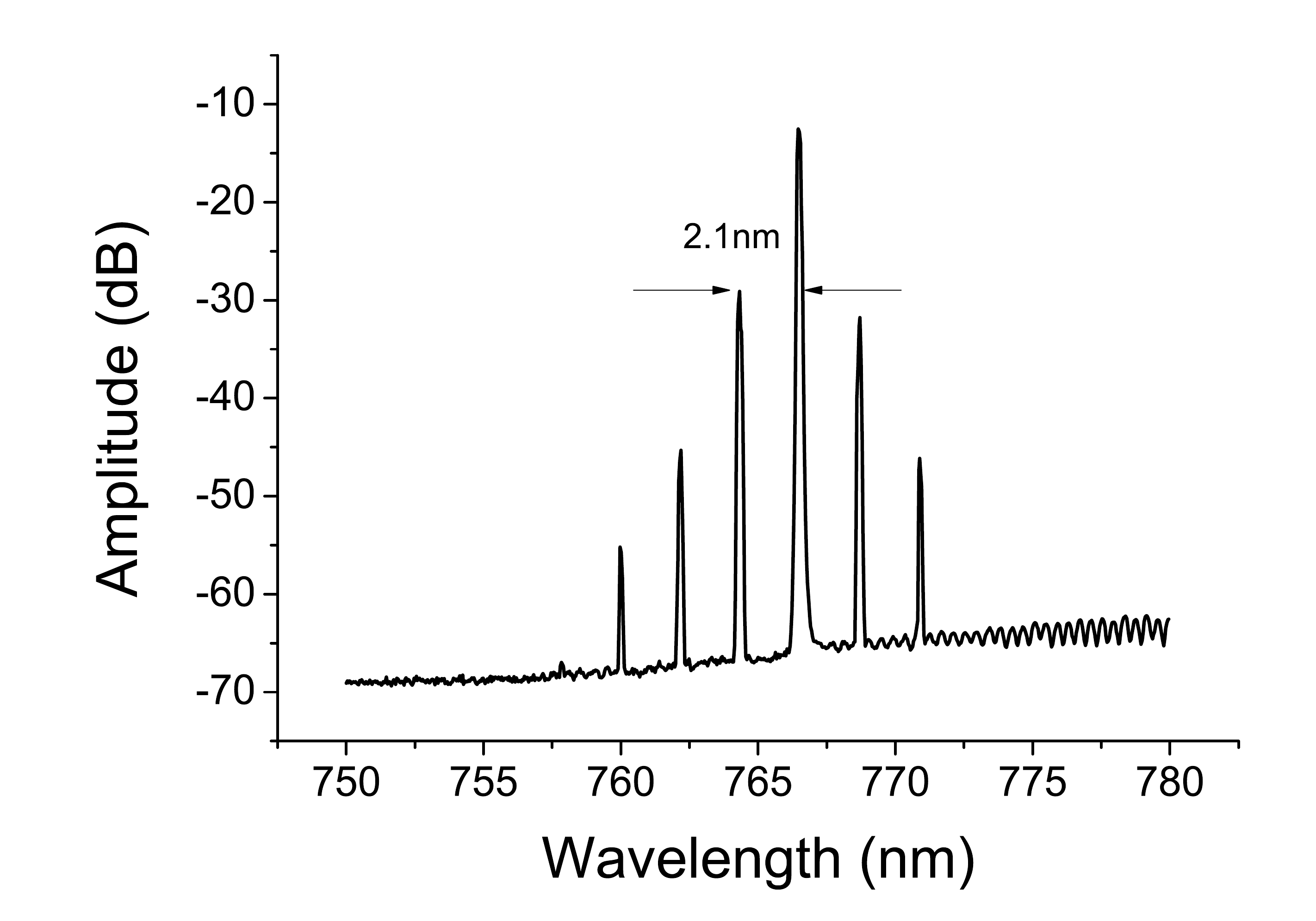}
\caption{\label{fig:fwm} Four-wave mixing spectrum of the MBR with a diameter of 120 $\mu$m, shown in Fig. \ref{fig:setup}(a)-(b). The separation between the peaks is 2.1 nm.}
\end{figure}

In the measurement, the pump laser was scanned over 40 GHz around a high \textit{Q} mode, and the transmission spectrum was recorded by a digital oscilloscope (Tektronics TDS 3014B). During the experiment, the coupling fiber was in contact with the MBR at the equatorial plane to maintain the coupling stability. A typical transmission spectrum and a resonance mode with a \textit{Q}-factor of $\sim 2\times 10^7$ are shown in Fig. \ref{fig:trans}. The span of the laser scan was then decreased until only the selected high \textit{Q }mode was coupled to the MBR. As the simulation results indicate, TE and TM modes in the MBR have different ZDWs. To effectively excite the TM mode, a fiber polarization controller (FPC) was used between the laser source and the tapered fiber. A typical FWM spectrum is presented in Fig. \ref{fig:fwm} with an input power of $\sim$3 mW at a pump wavelength of 766.45 nm. The symmetric equidistant lines on either side of the pump (which is the highest peak on the spectrum) are separated by 2.1 nm, which is roughly twice the calculated FSR of 1.1 nm. No Raman scattering signals were observed in this case, thus Fig. \ref{fig:fwm} shows  solid evidence of  degenerate FWM, \textit{i.e.} the MBR is in  the anomalous dispersion regime.

\begin{figure}
\centering
\includegraphics[width=.9\columnwidth]{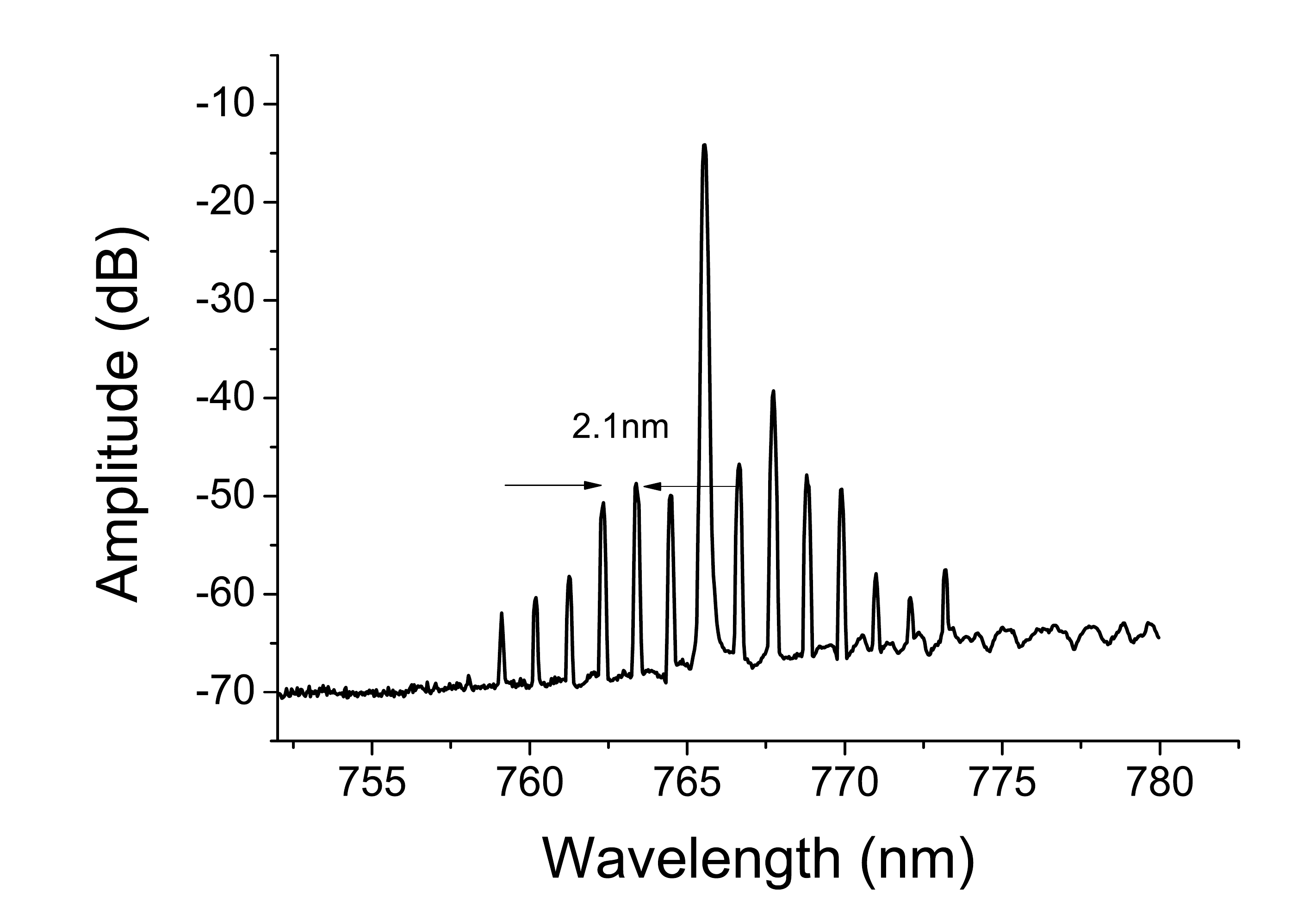}
\caption{\label{fig:fc} Frequency comb generation in the MBR at a center wavelength around 765 nm. Up to 14 comb lines are excited.}
\end{figure}

Next, we kept increasing the input power of the pump laser, and  more modes were excited via the FWM process until parametric oscillation occurred. Fig. \ref{fig:fc} shows the resulting spectrum at a power level of around 6 mW. Here, 14 peaks are visible and the separation between adjacent lines is also 2.1 nm. In this case, a 'Type I' (natively mode spaced) comb has been generated \cite{Herr2012}.

In summary, we have generated a frequency comb in a silica MBR by engineering the dispersion through optimizing the wall thickness and diameter of the MBR in a controllable way. The ZDW can be shifted beyond the limitation of the material's dispersion. We experimentally demonstrated FWM at 765 nm as a proof of principle. A frequency comb with multiple equidistant lines was generated by increasing the pump power to enter the parametric oscillation regime. The frequency comb generation was limited by the total available input power. The spectral bandwidth of the frequency comb increased with a higher power. In order to improve the frequency span of the comb, careful control of the dispersion is required; this could be implemented by introducing a small amount of curvature to the MBR. In practice, the wall thickness can be further decreased while still maintaining a high \textit{Q}-factor \cite{Yang:16}. Thence, the center wavelength of the frequency comb could be shifted to an even shorter wavelength, until it is eventually limited by the material absorption window. This mechanism of dispersion engineering can be used for other materials \cite{Riesen2016}, in particular, glass materials with high nonlinearity \cite{Wang2015}. In the future, an improved frequency comb in the visible wavelength range may be realized.

\section*{Acknowledgment}
This work was supported by the Okinawa Institute of Science and Technology Graduate University.
The authors would like to thank R. Madugani for technical assistance.
%

\end{document}